\documentclass[10pt]{article}

\usepackage{caption}
 \usepackage{subcaption}
 \usepackage{amsmath}
 \usepackage{graphicx}

\usepackage{color}

\newcommand{\bm}[1]{{\bf #1}}
\newcommand*{\bms}[1]{{\boldsymbol{#1}}}
\newcommand{\C}{\bm{C}}
\newcommand{\dv}{\bm{d}}
\newcommand{\F} {\bm{F}}
\newcommand{\G} {\bm{G}}
\newcommand{\I} {\bm{I}}
\newcommand{\M} {\bm{M}}
\newcommand{\T} {\bm{T}}
\newcommand{\p}{\bm{p}}
\newcommand{\vv}{\bm{v}}
\newcommand{\piv}{\bms{\pi}}
\newcommand{\rhov}{\bms{\rho}}
\title{Physical Limitations of Work Extraction from Temporal Correlations \\ }

\author{Elan Stopnitzky,\thanks{
{Department of Physics and Astronomy, 2505 Correa Rd, Honolulu HI 96822; elanstop@hawaii.edu}} * Susanne Still \thanks{Information and Computer Sciences, 1680 East-West Rd, Honolulu, HI 96822 and Department of Physics and Astronomy, Honolulu HI 96822; sstill@hawaii.edu}, Thomas E. Ouldridge \thanks{Department of Bioengineering and Imperial College Centre for Synthetic Biology, Imperial College London, London, SW7 2AZ, UK}, \\
and Lee Altenberg \thanks{Information and Computer Sciences, 1680 East-West Rd., Honolulu, HI 96822} \\ {\small *To whom correspondence should be addressed.}}

\date{} 

\begin{document}


\maketitle


\begin{abstract}
Recently proposed information-exploiting systems designed to extract work from a single heat bath utilize temporal correlations on an input tape. We study how enforcing time-continuous dynamics, which is necessary to ensure the device is physically realizable, constrains possible designs and drastically diminishes efficiency. We show that these problems can be circumvented by means of applying an external, time-varying protocol. This turns the device from a ``passive", free-running machine into an ``actively" driven one.
\end{abstract}


\section{Introduction}

Leo Szilard proposed a simple Gedankenexperiment almost 90 years ago to resolve the paradox of Maxwell's demon, arguing that information about a system could be converted to work by an automated mechanism, in place of a sentient being \cite{og_szilard}. Szilard's proposed information engine cyclically repeats two distinct phases: that of acquiring information and recording it into a stable memory, and that of using this information to extract work with a given mechanism. This allowed him to compute a bound on the costs associated with acquiring and recording information, necessary to prevent a violation of the Second Law \cite{og_szilard}. Many extensions to Szilard's engine have been explored in the literature, e.g. \cite{zurek,kim,marathe,sagawa,vaik}, and modern formulations of non-equilibrium thermodynamics naturally incorporate correlations as a potential source of work (see e.g. \cite{esposito2011}).

Much recent interest has nonetheless focused on the information utilization side, by building on the idea of a device that exploits a data-carrying tape  to extract work from a single heat bath \cite{jar}. Such a device advances along a sequence of 0s and 1s that contains an overall bias towards either 0 or 1. The device couples to one input bit at a time, and, whilst in contact with the bit, undergoes free-running dynamics that can alter the bit. This interaction increases the entropy of the tape upon output of the changed bit, and it is this entropy increase that is used to compensate for the entropy decrease of the heat bath. Extensions of these machines exploit statistical information within the tape in the form of temporal correlations \cite{jim}, or spatial correlations between tapes \cite{tom}, rather than an overall bias in the input bits. The resulting simple dynamical models of all these proposals help develop a concrete physical understanding of the role information plays in thermodynamics. To serve this purpose, it is important that these devices are physically realizable. 

Real physical systems have underlying time-continuous dynamics. {Moreover, whenever the work extraction device is designed to operate without a time-dependent, externally-applied driving protocol during the periods of interaction with an individual bit, then the time-continuous dynamics must also be time-homogeneous and obey detailed balance to be physical.}

We explore here how this fact constrains possible designs of the class of {\em temporal} correlation powered devices proposed in \cite{jim}, and some references therein. We find that demanding underlying time-continuous, time-homogeneous dynamics drastically limits the set of allowable transition matrices, thereby dramatically reducing the resulting efficiency (Section \ref{results}). {This fact is demonstrated first through the relative performance of randomly generated transition matrices and secondly via an evolutionary algorithm; in both cases we compare the situation with and without the constraint. We explain the difference in performance by the mathematical properties of the relevant matrices and their associated physical implications. Finally, we show that these limitations} disappear when the restriction to time-homogeneous dynamics is lifted and transition rates are modulated by external manipulation (Section \ref{sidestep}). 

\subsection{Model of a temporal correlation powered work extraction device.}\label{model} 
This model largely follows \cite{jim}. Imagine a work extraction system with two internal states, $s \in \{A,B\}$, which can be coupled to and decoupled from a work reservoir (such as a weight), an input tape with bits $b^{\it in} \in \{0,1\}$, an output tape with $b^{\it out} \in \{0,1\}$, and a heat bath.   The joint state-input value of the coupled system is then {$(s,b) \in \{A,B\} \times  \{0,1\} = \{A0,A1,B0,B1\}$, where $b$ denotes a coupled bit}.  Each of these four joint states possesses a potential energy, $E_i$, $i \in \{A0,A1,B0,B1\}$. The dynamics are described by a time-dependent vector containing the probabilities that the system is in one of the joint states at a given time,
$\p_{sb} \equiv [p_{A0},p_{A1},p_{B0},p_{B1}]^{\top}$ 
($^{\top}$ denotes the transpose). 

The engine alternates between an ``interaction step'', during which the internal state interacts with a bit, and a ``switching step'', during which the bit is changed.  Any changes in energy during an interaction step are due to the exchange of heat with the heat bath, and changes in energy during a switching step are due to exchange of work with the work reservoir. We therefore talk about ``heat" steps and ``work" steps, and will use these words to label transformations, as a reminder.

An interaction step is represented by the transformation $\p_{sb} \xrightarrow{\rm heat} \M \p_{sb}$, where the joint state evolves under the action of a reversible, column stochastic matrix, $\M$, for a duration of time, $\tau$.  During interaction steps, the system undergoes a free-running relaxation towards equilibrium. Thus, $\M$ must be reversible, and 
the stationary distribution, $\rhov^{(\M)} = \M \rhov^{(\M)}$, satisfies detailed balance

\begin{equation}\label{eq:detailed_balance}
M_{ij}\rho^{(\M)}_{j} = M_{ji}\rho^{(\M)}_{i} \quad\text{for all\ }  i, j \in \{A0,A1,B0,B1\}.
\end{equation} 

\noindent The coupled device and bit relax towards thermodynamic equilibrium, described by the Boltzmann distribution $\rho^{(\M)}_{i} = e^{-E_{i}/kT}/Z$. For simplicity, we choose units such that kT=1 and set the energy scale so that Z=1. We can then write the energy of each joint state as $E_{i} = -\ln \rho^{(\M)}_{i}$, with $i \in \{A0,A1,B0,B1\}$.

In a switching step, the internal state is held fixed, and a new bit is coupled to the device.
Whichever bit comprised the joint state prior to the switching step is printed to the output tape. That is, switching from the machine's $n$-th cycle to cycle $n+1$ changes the state of the bit that is interacting with the machine from $b = b_{n}^{out}$ to $b = b_{n+1}^{in}$. 
Depending on whether the input is 0 or 1, switching corresponds to the following transformation of the joint state probability vector:
\begin{eqnarray}
\p_{sb} = [p_{A0},p_{A1},p_{B0},p_{B1}]^{\top} & \xrightarrow{\rm input \; 0} &\bar{\p}_{sb} = [p_{A0}{+}p_{A1},0,p_{B0}{+}p_{B1},0]^{\top} \!\equiv\! \F_{0} \p_{sb}, \;\;\;\;\; \\
\p_{sb} = [p_{A0},p_{A1},p_{B0},p_{B1}]^{\top} &\!\!\!\!\!\!\!\xrightarrow{\rm input \; 1} \!\!\!\!\!\!\!&\bar{\p}_{sb}=[0,p_{A0}{+}p_{A1},0,p_{B0}{+}p_{B1}]^{\top} \!\equiv\! \F_{1} \p_{sb},  \;\;\;\;\; 
\end{eqnarray}
where the matrices $\F_{0}$ and $\F_{1}$ represent the switching:
\begin{align*}
\F_{0}&=\begin{bmatrix}
1&1&0&0\\
0&0&0&0\\
0&0&1&1\\
0&0&0&0
\end{bmatrix},
\qquad\text{and} \qquad
\F_{1}=\begin{bmatrix}
0&0&0&0\\
1&1&0&0\\
0&0&0&0\\
0&0&1&1
\end{bmatrix}.
\end{align*}

\noindent To extract work, the machine must on average raise the energy of the joint $(s,b)$-state during interaction steps (i.e.\ absorb heat), and lower the energy during switching steps (i.e.\ deposit energy into the work reservoir).

In the following, we limit our analysis for simplicity to an input tape consisting of alternating $1$s and $0$s. This is an interesting special case, because the per-symbol entropy of the input tape is maximal, as ${\sf Prob}(b^{\it in}\!=\!0)={\sf Prob}(b^{\it in}\!=\!1)\!=\!\frac{1}{2}$, and hence cannot be leveraged for work extraction. Any net gain is thus due to exploiting temporal correlations.

A single complete cycle of operation is defined by the product of transition matrices $\C = \M \F_{0} \M \F_{1}$ (reflecting the alternating switching and interaction steps), taking the probability distribution of the four states from $\p_{sb}^{(n)}$ to $\p_{sb}^{(n+2)} = \C \p_{sb}^{(n)}$.  
We require that repeated application of the matrix $\C$ to any starting distribution $\p_{sb}^{(0)}$ converges to a steady state distribution $\piv_0(s,b)$, defined by $\C\piv_0(s,b) = \piv_0(s,b)$.  Thus $\C$ must be a primitive matrix (irreducible and aperiodic), which is assured if $M_{ij} > 0$ for all $i,j \in \{A0,A1,B0,B1\}$. We can define a steady state distribution if we census the system at each step of the cycle ($\piv_1$ to $\piv_3$ in the equation array below). Starting with feeding in a 1, a cycle is then characterized by the following changes ({we use $b'$ to denote a bit that is about to be transferred to the output tape}):

\begin{eqnarray}
\label{dyn1}
\piv_0(s_{n-1},b_{n-1}') &\xrightarrow{\rm work}& \piv_1(s_{n-1}, b_{n}) = \F_{1}\piv_0(s_{n-1},b_{n-1}') \\
\label{dyn2}
\piv_1(s_{n-1},b_{n}) &\xrightarrow{\rm heat}&  \piv_2(s_{n},b_{n}') = \M \piv_1(s_{n-1},b_{n}) \\
\label{dyn3}
\piv_2(s_{n},b_{n}') &\xrightarrow{\rm work}& \piv_3(s_{n},b_{n+1}) = \F_{0} \piv_2(s_{n},b_{n}')  \\
\label{dyn4}
\piv_3(s_{n},b_{n+1}) &\xrightarrow{\rm heat}& \piv_0(s_{n+1},b_{n+1}') = \M \piv_3(s_{n},b_{n+1}) 
\end{eqnarray} 
The average work supplied to the work reservoir per input symbol is given by the sum of the average energy changes in the two switching steps:
$\langle W \rangle = - \frac{1}{2}\big[ \langle E \rangle_{\piv_0(s_{n-1},b_{n-1}')} - \langle E \rangle_{\piv_1(s_{n-1},b_{n})} + \langle E \rangle_{\piv_2(s_{n},b_{n}')} - \langle E \rangle_{\piv_3(s_{n},b_{n+1})} \big]$. The factor of $1/2$ is due to two bits being encountered per cycle.\footnote{
Note that this is work {\em extracted} from the joint ($s,b$) system. By convention, work done on the system is positive, as is heat flowing into the system.}

\section{Work extraction by time-continuous, free-running devices}\label{results}
We now depart from the approach of \cite{jim}. To be physically realizable in the absence of externally applied driving during the interaction period, the matrix $\M$ should correspond with a continuous-time equilibration process for some time $\tau$. In other words, we require there be a generator, $\G$, such that $\M=e^{\tau \G}$, where $\G$ is a reversible rate matrix with non-negative off-diagonal elements, in which every column sums to $0$. If $\M$ can be constructed in this way, then $\M$ is said to be ``reversibly embeddable'' \cite{jia}. The following results from \cite{jia} are crucial: 1) If $\M$ is reversible, then $\M$ is diagonalizable and the eigenvalues of $\M$ are real. 2) If $\M$ is also embeddable, then the eigenvalues of $\M$ are all positive, and the generator, $\G$, of $\M$ is unique.

\begin{figure}[ht]
\centering
\begin{subfigure}{0.5\textwidth}
 \centering
 \includegraphics[width=1\linewidth]{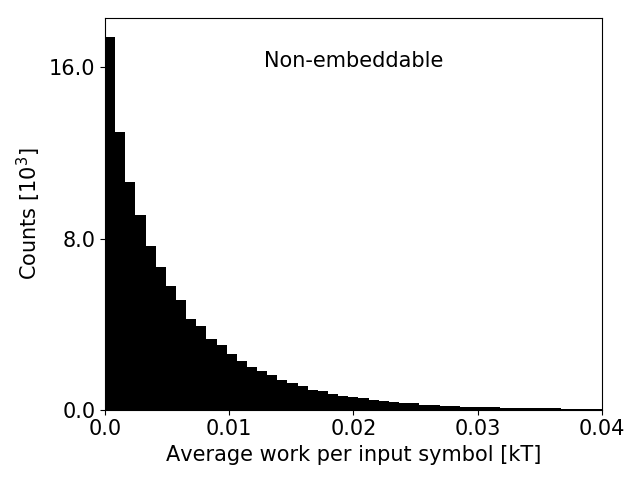}
 \label{fig1:sub1}
\end{subfigure}
\begin{subfigure}{0.5\textwidth}
 \centering
 \includegraphics[width=1\linewidth]{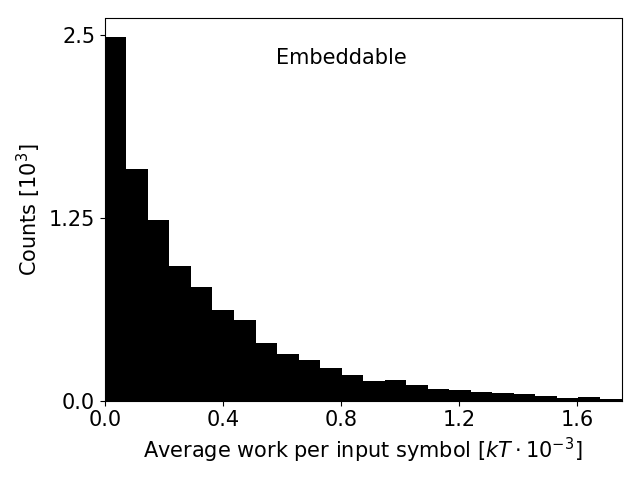}
 \label{fig1:sub2}
\end{subfigure}
\caption{Histogram of work, $W_{\rm out}$, extracted by randomly generated reversible transition matrices. Only positive work values are shown. With $10^{6}$ randomly generated matrices of each type, $11\%$ of non-embeddable matrices and $0.3\%$ of embeddable ones achieved positive work production.}
\label{fig:hist}
\end{figure}

Due to these properties, processes governed by reversible and embeddable transition matrices generally extract much less work than those governed by reversible but not embeddable ones, as we will see shortly. In the rest of this paper, we will only be considering matrices that are reversible. For brevity, we henceforth use the terms {\it embeddable} and {\it non-embeddable} to refer to the two different classes of matrices. 
In Figure \ref{fig:hist}, we display histograms showing the number of randomly generated matrices that achieve various values of positive work, for the two categories. The procedure used to make these matrices is detailed in the Appendix. The best randomly found embeddable matrices extract roughly a factor 20 less work than the best randomly found non-embeddable ones. In comparison, the construction given in Figure 6 of \cite{jim}, which is non-embeddable, extracts $\frac{kT}{e} \approx 0.368 \;kT$ of work per input symbol. We see from the histogram that finding a machine with a comparable efficiency by chance is rather unlikely.

\begin{figure}[ht]
\centering
 \includegraphics[width=0.5\linewidth]{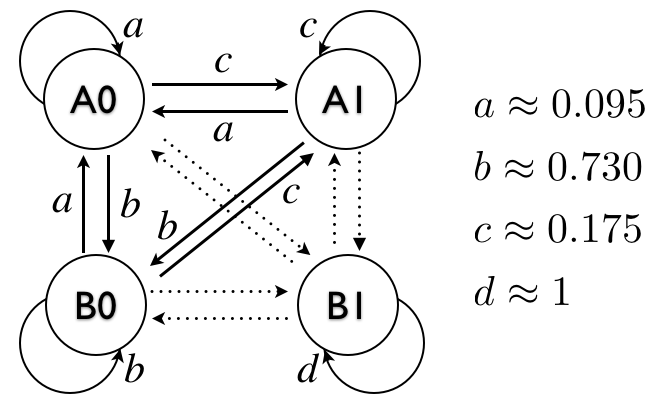}
 \caption{Design of the best embeddable transition matrix found by the evolutionary algorithm, with $W_{\it out} \approx 0.0146$ $kT$. Dotted arrows denote transition probabilities close to zero. The transition matrix for this graph has a rank-1 submatrix, a second eigenvalue very close to 1, and the two smallest eigenvalues very close to zero.}
\label{fig:drawing} 
\end{figure}

To improve upon random sampling, we constructed an evolutionary algorithm to explore the search space. The algorithm applied mutations to individual machine dynamics and fixed those mutations whenever they led to improved performance. Due to the high dimension of the space of transition matrices, this algorithm performed better than a grid search. When ignoring the embeddability constraint, the evolutionary algorithm readily returned the best design of \cite{jim}, and never found one better. However, enforcing embeddability lowered the efficiency drastically. The best embeddable design which the evolutionary algorithm found is shown in Figure \ref{fig:drawing}. It achieved only $\approx$ 0.0146 $kT$ per input symbol. That is less than 4\% of the yield of the non-embeddable best design of \cite{jim}.
 
We now discuss why the performance of embeddable designs is so poor. Optimal performance requires that the internal state of the device, $s_n$, contain predictive information\footnote{We detail in the Appendix how to calculate information.}, $I(s_{n};b_{n+1})$, about the next incoming bit, $b_{n+1}$. To see why, note that the non-equilibrium free energy of system coupled to bit, $F[\pi] = \langle E \rangle_\pi - k T H[\pi]$, cannot increase spontaneously during an interaction step. 
Thereby, the heat absorbed in one interaction step is upper bounded by the entropy change, which can be written as
$\langle Q_n \rangle \leq kT\big[I(s_{n};b_{n+1})-I(s_{n+1};b_{n+1}')+H(s_{n+1})-H(s_{n})+H(b_{n+1}')-H(b_{n+1})\big]$.
Adding two of these heat contributions to account for the full cycle of Eqs. (\ref{dyn1}-\ref{dyn4}), we get a cancelation, because $H(s_{n+1})-H(s_{n-1}) = 0$, {due to the fact that we are in the same stationary distribution $\piv_0$ at the beginning and end of the cycle.} 

\begin{figure}[ht]
\centering
\includegraphics[scale=0.5]{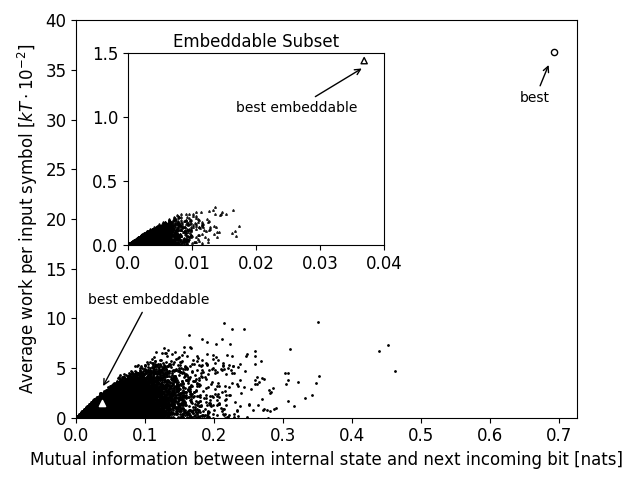}
\caption{Average work per input symbol, $W_{\it out}$, vs. average predictive information per symbol. The triangle denotes the best embeddable design found by the evolutionary algorithm. The circle denotes the design of \cite{jim}. The inset shows only the subset of embeddable designs. See Appendix for further details. }
\label{fig:info}
\end{figure}

Taking the average then sets an upper bound on the extractable work per input symbol, $W_{\rm out}$. The bound depends on how the average predictive information about the input, 
{$I_{\rm pred} = (I(s_{n-1};b_{n}) + I(s_{n};b_{n+1}))/2$} compares to the average memory about the output, \mbox{$I_{\rm mem} = (I(s_{n};b_{n}') + I(s_{n+1};b_{n+1}'))/2$}, and on how the average output entropy, $H_B^{\rm out} = (H(b_{n}')+H(b_{n+1}'))/2$ compares to the average input entropy, \mbox{$H_B^{\rm in} = (H(b_{n})+H(b_{n+1}))/2$}:
\begin{equation}
W_{\rm out} \leq kT\big[I_{\rm pred} - I_{\rm mem} +\Delta H_B\big]~,
\end{equation}
where $\Delta H_B \equiv H_B^{\rm out} - H_B^{\rm in}$. 
We display $W_{\rm out}$ as a function of $I_{\rm pred}$ in Figure \ref{fig:info}, for each of the two classes of transition matrices. It is clear from the plot that predictive information between device and next incoming bit is severely limited for embeddable systems, and that there is a consequent reduction in extractable work. 

Perfect prediction of the next incoming bit requires synchronization with the input. For a period-2 input, the internal state must change in each interaction step from $A$ to $B$ or vice versa, necessitating a bipartite graph structure, whose associated transition matrix has negative eigenvalues \cite{gallager}, and therefore is non-embeddable. Tracking a periodic signal with period greater than $2$ would require complex eigenvalues \cite{gallager}, and is therefore impossible for any reversible matrix, embeddable or otherwise. 

\begin{figure}[ht]
\centering
\begin{subfigure}{0.5\textwidth}
 \centering
 \includegraphics[width=1\linewidth]{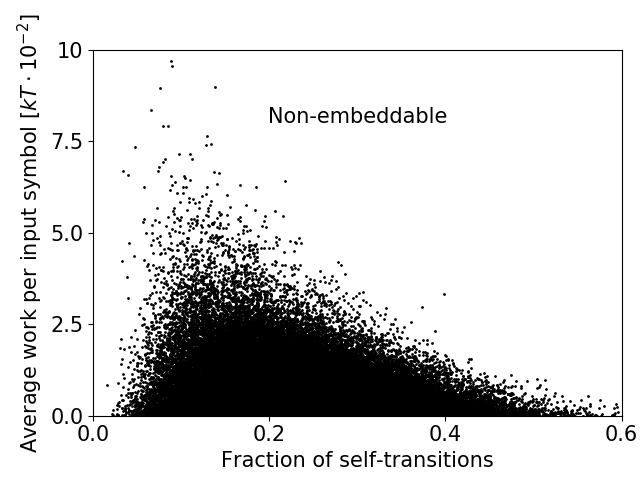}
 \label{fig4:sub1}
\end{subfigure}
\begin{subfigure}{0.5\textwidth}
 \centering
 \includegraphics[width=1\linewidth]{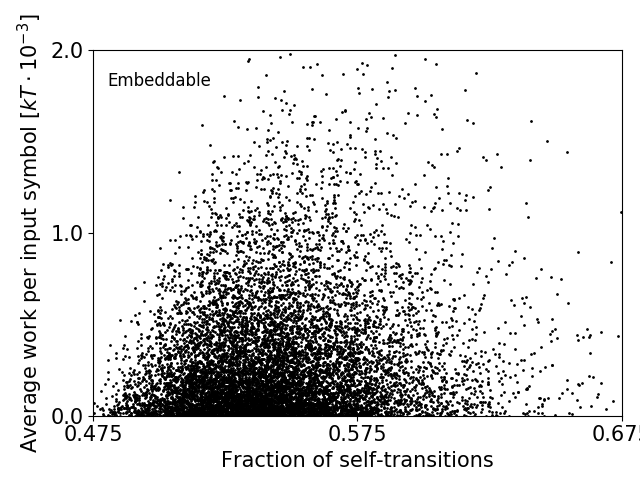}
 \label{fig4:sub2}
\end{subfigure}
\caption{The relationship between $W_{\it out}$ and the fraction of self-transitions}
\label{fig:self}
\end{figure}

Synchronization is hampered by the tendency of embeddable systems to undergo ``self-transitions'', in which the system starts and ends in the {\it same} state during an interaction interval. These self-transitions are also undesirable because they are associated with no net exchange of energy with the bath, thus wasting the input. Self-transitions arise from the diagonal entries in $\M$, which can be set to zero for non-embeddable $\M$ \cite{jim}, but not if $\M$ is embeddable. To see why, note that $\M$ being a stochastic matrix implies that it has an eigenvalue of $1$, and $\M$ being embeddable implies that all other eigenvalues are positive. Thus, the trace of $\M$ must be greater than $1$ and self-transitions cannot be neglected. 
The average fraction of such self-transitions in interaction steps can be written as
\begin{equation}\label{eq:selftrans}
\frac{1}{2} \dv(\M)^{\top} [\piv(s_{n},b_{n}) + \piv(s_{n+1},b_{n+1}) ],
\end{equation}
where $\dv(\M)$ is a vector of the diagonal elements of $\M$.
We show in the Appendix that for the setup considered here, self-transitions occur at a minimum of $1/4$ of interaction steps. Note that this feature implies that there do not exist any embeddable matrices ``close'' to the non-embeddable design of \cite{jim}. The relationship between self-transition rate and average work extracted for randomly chosen embeddable and non-embeddable designs is shown in Figure \ref{fig:self}. 
We display only those designs that lead to positive work extraction. There are fewer points in the right panel because a smaller fraction of embeddable matrices lead to positive average work. Since the {non-embeddable} matrices can be made to have a trace of zero, their rate of self-transitions can also be made $0$, while for the {embeddable} matrices a minimum trace of $1$ forces the rate of self-transitions to be at least $1/4$, leading to inefficiency.
Moreover, approaching the minimal trace of $1$ and the corresponding minimal self-transition rate of $1/4$ is not a viable strategy for maximizing efficiency, because it would mean that all the smaller eigenvalues would have to approach zero. This requirement is in tension with the fact that the modulus of the second largest eigenvalue, $\lambda_2 < 1$, bounds the distance from equilibrium at the end of an interaction step through $\M\p = \rhov^{(\M)} + \lambda_{2} c_{2}\vv_{2}+\lambda_{3}c_{3}\vv_{3}+\lambda_{4}c_{4}\vv_{4}$, where $\p= \rhov^{(\M)} + c_{2}\vv_{2}+ c_{3}\vv_{3}+ c_{4}\vv_{4}$ is an arbitrary starting distribution expanded in the basis of $\M$'s eigenvectors, $\rhov^{(\M)}$,$\vv_{2}$,$\vv_{3}$,$\vv_{4}$. Thus, taking the limit as the trace approaches $1$ would cause instantaneous relaxation to equilibrium in every interaction step, which prevents work from being extracted.  

Intuitively, the device cannot extract work if it relaxes fully to equilibrium in each interaction step because in that case it cannot retain memory. More formally, let $E_{eq}$ denote the energy of the equilibrium distribution. With complete relaxation to equilibrium in each interaction step, the sum of the energy changes over the two switching steps is then ${1\over Z} [E_{A1}e^{-E_{A0}}+E_{A0}e^{-E_{A1}}+E_{B0}e^{-E_{B1}}+E_{B1}e^{-E_{B0}}] - E_{eq} \geq 0$. This quantity is non-negative because the equilibrium distribution pairs the highest energies with the smallest Boltzmann factors, so any reordering of the factors cannot decrease the average energy.

Altogether, there exists a tradeoff between inefficiency coming from staying in the same state too often, and inefficiency coming from relaxing too close to equilibrium.
This tradeoff would be less severe if more internal states were included, because with a transition matrix of higher dimension, the trace could be kept relatively small even with a large second eigenvalue. 
However, adding additional internal states would not necessarily guarantee a substantial improvement of the overall performance, because the prohibition on bipartite graphs prevents synchronization with the input, an issue that persists.

Much of the inefficiency of embeddable designs arises from an inability to reliably track the input by switching the device state at each step. But embeddable designs also suffer from a second drawback that would limit work extraction even if the input were a pure string of 1s. To extract work, it is vital that the energy tends to increase during the interaction window. The work extracted is equal to the average number of transitions during these windows, multiplied by the increase in energy per transition. But if we increase the energy of the high energy states to which we hope the system will transition, then we decrease the net number of transitions, since we decrease their occupancy in equilibrium -- and embeddable designs can only relax towards equilibrium during the interaction window. Thus there is an unavoidable trade-off for embeddable designs between facilitating many transitions that each contribute only a little to the work extracted, and allowing only a few that contribute a large amount. The overall power is maximised at intermediate values (see, for example, \cite{jar}).

\section{Time-Inhomogeneous Protocols}\label{sidestep}

We have shown that devices that are free-running during the interaction period, which are restricted to reversibly embeddable, time-homogeneous dynamics, can extract only a small fraction of the work available from an alternating input of 0s and 1s. One might ask whether devices connected  to a time-dependent, externally-applied protocol during the interaction period, resulting in time-inhomogeneous dynamics that need not satisfy detailed balance,
could perform better. In this setting, for example, it can be ensured that the device's state \textit{must} change during a cycle, allowing a better match to the input's periodicity. 

Here we show that it is relatively straightforward to construct a device of this type that extracts, in the quasistatic limit, all the work stored in an input tape of alternating 1s and 0s.
External manipulations correspond to changing the energy levels of the system over time \cite{christian}, and allow the input and/or extraction of work during the interaction period of duration $\tau$. For our purposes, it is sufficient to consider only devices in which the energies of the joint states ($s,b$) vary during $\tau$, but are restricted to all being {\it identical} at the beginning and end of each window. In this case, the work of switching to the next input bit on the tape is zero, and only the window $\tau$ need be considered to compute the extracted work.

To demonstrate an optimal device, let us compose it of two standard operations: erasure and relaxation. Let a given pair of states within a larger state space each have an occupation probability of $p/2$. Erasure shifts all of this probability to just one state, leaving it occupied with probability $p$, and the other with a probability 0. Famously, erasure can in principle be performed at a work cost of $p k_{\rm B}T \ln 2$ \cite{Landauer,Wolpert}. Relaxation is the inverse of erasure, and  therefore work of $p k_{\rm B}T \ln 2$ can be extracted. In both cases, these optimal limits on the work are reached by {\it thermodynamically} reversible processes, in which manipulations must be applied quasistatically and the reversal of the protocol would restore the initial probability distribution. We also consider switching, or the transfer of probability from one state to another that has zero initial probability. Switching can be decomposed as a relaxation followed by an erase, and therefore has no
 total work requirement if performed in a thermodynamically reversible manner.

Let us consider the following transition matrix for the interaction step
\begin{equation}
\T = \begin{bmatrix}
 0 & 1/2 & 0 & 1/2\\ 0 & 1/2 & 0 & 1/2 \\ 1/2 & 0 & 1/2 & 0 \\ 1/2 & 0 & 1/2 & 0 
\end{bmatrix},
\end{equation}

\noindent where $\T \p_{sb}$ gives the evolution of $\p_{sb}$ during a single interaction window. This transition matrix ensures that the machine transitions to the $A$ state if the input bit was in state 1, and to the $B$ state if the input bit was in state $0$.  This oscillation is the central switching motif that allows the device to track the input.  $\T$ is produced by composition of the following sequential operations: a switch from $(A,0)$ to $(B,0)$; a switch from $(B,1)$ to $(A,1)$; a relaxation from $(B,0)$ to $(B,0)$ or $(B,1)$; and finally a relaxation from $(A,1)$ to $(A,0)$ or $(A,1)$.  The states $(A,1)$ and $(B,0)$ are effectively used as ancillary states \cite{Wolpert} to facilitate the necessary reversing of the machine's state at each step, prior to relaxation. 

Regardless of the initial condition, a single application of $\F_1 \T \F_0 \T $ will bring the system to 
\begin{equation}  
\hat{\p}_{sb} = \big[p_{A0}, p_{A1}, p_{B0}, p_{B1}\big]^{\top} =  \big[0,0,0,1 \big]^{\top},
\label{eq:tom1}
\end{equation} 
in which the state of the device and tape are perfectly coordinated. Since $\hat{\p}_{sb}$  is an eigenvector of $\F_1 \T \F_0 \T $ with eigenvalue 1,  subsequent applications of $\F_1 \T \F_0 \T $ will also return $\hat{\p}_{sb}$. Given the initial condition of $\hat{\p}_{sb}$, the switch and relaxation procedures underlying $T$ can be implemented in a thermodynamically reversible manner, yielding $k_BT \ln2$ of work per $T$ operation (or $2k_BT \ln2$ per full cycle)  due to  the relaxation steps. Thus the device extracts all of the available work after an initial alignment cycle.

\section{Discussion and Conclusion}
We have learned that building physically realistic devices that exploit temporal correlations with a well-defined period to extract work from a heat bath at high efficiency can be challenging. Specifically, devices with time-continuous dynamics cannot extract much work from an alternating input of 0s and 1s, if they operate in a free-running fashion during interaction with the input bit. External manipulation by a time dependent protocol alleviates this issue. {We leave exploration of work extraction from inputs with a different temporal correlational structure for future work. However, as per the discussion in Section \ref{sidestep}, external manipulation is likely to be generally key for optimal work extraction, as it can guarantee that necessary transitions occur.}

For highly efficient {work-extraction} systems to emerge (for example due to an evolutionary process), they would then have to develop the ability to operate in an actively driven fashion, rather than passively, in order to reach maximum efficiency. An interesting implication that arises from this is the need for a higher order control structure for active driving. Hierarchical organizations are ubiquitous in biology, and it would be interesting to modify our evolutionary algorithm to explore the emergence of hierarchical structures for greater work extraction. 

For such systems, there will be a trade-off between the speed at which they operate and the amount of energy they can extract.
In our simple example, we can extract all the work we put in, plus extract net gain from the heat bath,
because external manipulations are applied reversibly. But, if we were to put constraints on the execution time, then we should see a trade-off between power and efficiency, similar to effects discussed for example in \cite{VdB16}, and references therein. We leave a systematic study of this effect to future work.

\section{Acknowledgements}
We would like to thank Jim Crutchfield, Alexander Boyd, Christopher Jarzynski, David Sivak, Rob Shaw, David Wolpert, Artemy Kolchinsky, Karoline Wiesner, and Michael Lachmann for relevant discussions. This work was partially supported by the Foundational Questions Institute (Grant No..\ FQXi-RFP3-1345).  LA received support from the University of Hawai`i Office of the Vice Chancellor for Research, the John Templeton Foundation, the Stanford Center for Computational, Evolutionary and Human Genomics, and the Morrison Institute for Population and Resources Studies, Stanford University.  TEO is supported by a Royal Society University Research Fellowship.  

\bibliographystyle{unsrt}
\bibliography{BitBeasts.bib}

\section{Appendix}

\subsection{Predictive information in steady state}

{In steady state, the distribution of internal states prior to receiving a $1$ is $\piv_{0,s} = [\pi_{0,A},\pi_{0,B}] = [\pi_{0,A0}+\pi_{0,A1},\pi_{0,B0}+\pi_{0,B1}]$ (see Eq. \ref{dyn1}). The distribution prior to receiving a $0$ is $\piv_{2,s} = [\pi_{2,A},\pi_{2,B}]=[\pi_{2,A0}+\pi_{2,A1},\pi_{2,B0}+\pi_{2,B1}]$ (see Eq. \ref{dyn3}). The overall probability of being in state $A$ at the end of an interaction step is then $\pi(A) = \frac{1}{2}[\pi_{0,A0}+\pi_{0,A1}+\pi_{2,A0}+\pi_{2,A1}]$, and $\pi(B) = 1-\pi(A)$. With these expressions, and noting that the overall probability of receiving each bit is $1/2$, the mutual information between the internal state at the end of an interaction interval and the next incoming bit simplifies to: $I_{pred} = \frac{1}{2}[\pi_{0,A} \ln \frac{\pi_{0,A}}{\pi(A)}+\pi_{0,B} \ln \frac{\pi_{0,B}}{\pi(B)}+\pi_{2,A} \ln \frac{\pi_{2,A}}{\pi(A)}+\pi_{2,B} \ln \frac{\pi_{2,B}}{\pi(B)}]$.} 

\subsection{Making reversible random matrices}

The following procedure for making reversible matrices at random is taken from \cite{random}. Let $\{K_{ij}| \, j \leq i \leq 4\}$ be a set of $10$ real, random variables created by sampling uniformly from an interval $(0,N_{max}]$. This set forms the lower triangle of a symmetric $4$ by $4$ matrix with $K_{ij} = K_{ji}$. Define $\pi_{j} \equiv \sum_{i}K_{ij}$. Then the matrix $\M$ given by $M_{ij} = \frac{K_{ij}}{\pi_{j}}$ is a reversible stochastic matrix, with stationary distribution $\piv$. We can also make generators $\G$ via $\G = \M-\I$, where $\I$ is the identity matrix. We generated $10^{6}$ $\M$ by this procedure with $N_{max} = 100$, as well as the corresponding generators given by $\G = \M-\I$. Note that this procedure gives transition rates in the range $(0,1]$. From this set of generators, we made $3\times10^{6}$ embeddable transition matrices, with $10^{6}$ each for $\tau=1$,$\tau=0.01$,and $\tau=100$. The interaction intervals with $\tau$ other than 1 led to poor performance, because the other values led to very high self-transition rates ($\tau=.01$) and full equilibration during the interaction interval ($\tau=100$). 

\subsection{Lower bound on self-transitions}

The overall chance of making a self-transition under the action of $\M$ on distribution $\p$ is 
$\dv(\M)^{\top} \p$,  where $\dv(\M)$ is the vector of the diagonal entries of $\M$. This quantity is minimized when the trace of $\M$ is minimal. For embeddable stochastic matrices, the lower bound on the trace is $1$ (see argument in Section \ref{results}). In this case, $\M$ has a single eigenvalue of $1$ and all other eigenvalues $0$. Thus, $\M$ is rank-1 with $4$ repeats of the same column ${\bf m} = [m_{0},m_{1},m_{2},m_{3}]^{\top}$. The matrix $\C = \M\F_{0}\M\F_{1}$ is then equal to $\M$, and the steady state of the complete cycle is nothing more than the repeated column, i.e., $\piv_0(s_{n},b_{n-1}') = {\bf m}$. Equation (\ref{eq:selftrans}) says that the average number of self-transitions over the cycle is $L = \frac{1}{2}\big[(m_{0}+m_{1})^{2}+(m_{2}+m_{3})^{2}\big]$. Minimizing this number is a simple optimization problem with the constraint $m_{0}+m_{1}+m_{2}+m_{3} = 1$. The solution is $(m_{0}+m_{1}) = (m_{2}+m_{3}) = \frac{1}{2}$. Substituting these in gives $L = \frac{1}{4}$. This condition can be satisfied for approximately embeddable $\M$, which can be constructed, for example, by perturbing the matrix with all entries equal to $1/4$ so that the smaller eigenvalues are just slightly positive and not exactly zero.

\end{document}